\documentclass[aps,prev,twocolumn,preprintnumbers,floatfix,nofootinbib]{revtex4-1}
\usepackage{graphicx}
\usepackage{bm}
\usepackage{times}
\usepackage{slashed}
\usepackage{color}
\usepackage{color}
\pdfoutput=1

\begin{document}

\title{Non-thermal WIMPs as ``Dark Radiation'' in Light of ATACAMA, SPT, WMAP9 and Planck}

\author{Chris Kelso$^{a}$}
\author{Stefano Profumo$^{b}$}
\author{Farinaldo S. Queiroz$^{b}$}

\affiliation{$^{a}$Department of Physics and Astronomy,
University of Utah, Salt Lake City, UT 84112, USA.
\\
$^b$ Department of Physics and Santa Cruz Institute for Particle Physics
University of California, Santa Cruz, CA 95064, USA
}

\date{\today}
\vspace{1cm}

\begin{abstract}

The Planck and WMAP9 satellites, as well as the ATACAMA and South Pole telescopes, have recently presented  results on the angular power spectrum of the comic microwave background. Data tentatively point to the existence of an extra radiation component in the early universe. Here, we show that this extra component can be mimicked by ordinary WIMP dark matter particles whose majority is cold, but with a small fraction being non-thermally produced in a relativistic state. We present a few example theories where this scenario is explicitly realized, and explore the relevant parameter space consistent with BBN, CMB and Structure Formation bounds.
\end{abstract}

\maketitle

\section{INTRODUCTION}
\label{introduction}

The nature of dark matter (DM) is one
of the major mysteries in our current understanding of the universe. A comprehensive strategy encompassing direct and indirect detection plus collider searches, as well as indirect information from cosmological observations, will certainly be necessary to determine the fundamental nature of the DM. The results from the DAMA/LIBRA Collaboration~\cite{Bernabei:2013qx} and, more recently, from the CoGeNT~\cite{Aalseth:2010vx,Aalseth:2011wp}, CRESST~\cite{Angloher:2011uu} and CDMS~\cite{Agnese:2013rvf} experiments, might point to the first direct signals of light Weakly Interacting Massive Particles (WIMPs) scattering off of nuclei \cite{Chris1}. 
The gamma-ray excess from the Galactic Center region \cite{GCexcess} and the tentative 
gamma-ray line observed with the Fermi Large Area Telescope \cite{Gammaline} have also been interpreted as resulting from WIMP annihilation. The excess cosmic-ray positrons, recently confirmed by AMS-02 \cite{ams02}, is an additional puzzling result that some have associated with WIMP annihilation~\cite{Kopp:2013eka,DeSimone:2013fia,Ibe:2013nka,Jin:2013nta} (see however \cite{Linden:2013mqa}).

Besides these putative direct and indirect detection signals, recent measurements of 
the angular power spectrum of the cosmic microwave background (CMB) by a variety of telescopes and 
satellites seem to indicate the existence of an extra component of radiation in the early universe. In particular, the WMAP Collaboration has recently presented their 9-year data and concluded that, after combining with data from Baryonic Acoustic Oscillations (BAO), from the Atacama Cosmology Telescope (ACT), and from the Hubble Space Telescope (HST) ($ H_0 = 73.8\ \pm  2.4\ {\rm km\ s}^{-1}\ {\rm Mpc}^{-1}  $, \cite{HST}), the number of effective massless neutrinos is $N_{eff} = 3.84 \pm 0.4$ \cite{WMAP9}.

The situation is, however, quite complex and subtle when the results of other experiments are considered. For example, the SPT collaboration has also recently reported new results. Combining their data with BAO and HST, they concluded that $N_{eff} = 3.71 \pm 0.35$ \cite{SPT}. Depending on the data set used, these conclusions can, however, qualitatively change. For instance, combining data from WMAP9 and SPT, Ref.~\cite{Globalanalyses} finds $N_{eff} = 3.93 \pm 0.68$, while from a combination of WMAP9, SPT, and the 3-year Supernova Legacy Survey (SNLS3) \cite{Supernova}, $N_{eff} = 3.96 \pm 0.69$ \cite{Globalanalyses}. 
ACT also  reported recently a much lower value than what quoted above, namely $N_{eff} = 2.79 \pm 0.56$ \cite{ACT}, when utilizing WMAP7 data, and $N_{eff} = 2.74 \pm 0.47$ when including WMAP9 data \cite{Globalanalyses}. This value can, however, be much larger if one includes BAO and HST data, reaching $N_{eff} = 3.43 \pm 0.36$ \cite{Globalanalyses}. It is clear that SPT and ACT results are somewhat in tension, and this may be related to the different values used for the lensing amplitude parameter $A_L$ \cite{SPTACT}.

Lastly, the eagerly awaited Planck results have just been reported. At face value, Planck data alone seem to disfavor an extra radiation component, pointing to $N_{eff} = 3.36 \pm 0.34$ \cite{Planck}. It is important to note, however, that the Planck collaboration adopted fairly low values (at the $2.2\sigma$ level) for $H_0$ compared to previous studies \cite{HST}. Since $N_{eff}$ and $H_0$ are positively correlated, increasing $H_0$ would naturally yield higher values for $N_{eff}$ \cite{Planck}. Interestingly, adding the $H_0$ measurement and BAO data, the Planck Collaboration  finds, in fact, $N_{eff} = 3.52 ^{+0.48}_{-0.45}$ \cite{Planck}. This latter value is $\sim1\sigma$ above the $N_{eff}=3.046$ value favored by Big Bang Nucleosynthesis (BBN) \cite{BBN3}.

As the discussion above implies, the best fit value for $N_{eff}$ varies significantly depending on the data set used. Regardless of that, the combined data reviewed above seem to be tentatively pointing towards the existence of an extra radiation component. Interestingly, it has been shown that this additional component can be interpreted as a thermally produced light DM 
species \cite{CMBDMspecies}, or as {\em non-thermally} produced {\em relativistic} DM particles, as shown 
in a model-independent way in Ref.~\cite{Neffpaper}, and in the context of 
supersymmetric frameworks in Ref.~\cite{CMBDMsusy}. 

In this article, we present a few illustrative models, including a supersymmetric example, where the tentative extra radiation component is interpreted in terms of WIMP DM primarily composed of a standard cold component plus a small fraction (roughly $1\%$ or less) produced non-thermally in a relativistic state.  In Section~\ref{sec:derive} we relate the energy density associated with non-thermally, relativistically produced DM with the number of effective neutrinos. In Section~\ref{sec:cosmo} we examine the cosmological bounds from BBN, structure formation and the CMB on this scenario.  In Section~\ref{sec:models} we present four models that satify the cosmological bounds and produce $\Delta N_{eff} \simeq 1$. We summarize and conclude in Section~\ref{sec:conclusion}.

\section{LIGHT DARK MATTER MIMICKING AN EXTRA NEUTRINO IN THE EARLY UNIVERSE}
\label{sec:derive}
In order to relate the energy density associated with non-thermally, relativistically produced DM with the number of effective neutrinos, we start by calculating the ratio between their respective energy densities.  Since the Cold Dark Matter (CDM) and neutrino densities are given by $\rho_{DM} = \rho_c \Omega_{DM} a^{-3}$ and $\rho_{\nu} = \rho_c \Omega_{\nu} a_{eq}^{-4} N_{\nu}/3$, at Matter-Radiation Equality (MRE), the ratio between neutrino and DM energy density is 

\begin{equation}
\frac{\rho_{\nu}}{\rho_{DM}}= \frac{\Omega_{\nu}}{\Omega_{DM}}\ \frac{N_{\nu}}{3}\ \frac{1}{a_{eq}}=\frac{0.69\ \Omega_{\gamma}}{\Omega_{DM}} \frac{N_{\nu}}{3} \frac{1}{a_{eq}},
\label{rhoDMnu}
\end{equation}where $\Omega_{\gamma} \simeq 4.84 \times 10^{-5}$, $\Omega_{DM} \sim 0.227$, $N_{\nu}$ is the number of neutrinos, and $a_{EQ} \simeq 3\times 10^{-4}$ is the scale factor at MRE.

For $N_{\nu}=1$, we thus find that the energy density of one neutrino is $\sim16\%$ of the CDM density. As a result, if DM particles had a kinetic energy equivalent to $\gamma_{DM} \simeq 1.16$ at MRE, this fraction would produce the same effect as an extra neutrino species in the expansion of the universe at MRE, as already pointed out in Ref.~\cite{Neffpaper}.

Assuming that some fraction $f$ of the DM in the universe is produced via the decay $X^{\prime} \rightarrow DM + \gamma$, we can relate the boost factor of the DM particle produced in the decay process (see Appendix \ref{boostfactor}) with the number of additional effective neutrinos in the universe via

\begin{eqnarray}
\Delta N_{eff} & \simeq & 4.87\cdot10^{-3}\left( \frac{\tau}{10^6\ s} \right)^{1/2}\nonumber\\ 
                     &   &
  \times \left[\left( \frac{ M_{X^{\prime}} }{2M_{DM}} + \frac{M_{DM}}{2M_{X^{\prime}}} -1 \right) \right] f.
\label{deltaNeff}
\end{eqnarray}
Eq.~(\ref{deltaNeff}) illustrates that this scenario posseses three free parameters: 
\begin{itemize}
\item[(i)] the $X^\prime$ lifetime for the decay process, 
\item[(ii)] the mass ratio $M_{DM}/M_{X^\prime}$, and 
\item[(iii)] the fraction $f$ of the DM density produced via the decay. 
\end{itemize}
With specific choices for the 3 parameters, one can easily reproduce $\Delta N_{eff}^{\nu} \simeq 1$ and therefore mimic the effect of an extra effective relativistic degree of freedom \cite{Neffpaper}.  

It is important to point out that in the present setup most of the DM would still be produced thermally, for example via the standard thermal freeze-out picture, with only a small fraction being produced non-thermally and with a large kinetic energy. We show in the next section that the fraction of DM produced relativistically must be small ($f\ll1$) due to the fact this component directly affects the formation of structure and other cosmological bounds.  For example, the electromagnetic energy released in the decay could have catastrophic impacts on the abundance of light elements as synthesized in the Big Bang Nucleosynthesis (BBN) framework, as well as on the spectrum of the Cosmic Microwave Background radiation (CMB). We analyze in detail structure formation, BBN and CMB bounds on $f$ and on the $X^\prime$ lifetime in the following section.

\section{COSMOLOGICAL BOUNDS}
\label{sec:cosmo}
\subsection{Structure Formation}

It is well known that the primordial universe had small inhomogeneities in the density distribution of matter, which evolved linearly at first, while eventually growing non-linearly to form the structures observed today. In the standard picture of structure formation, the dark matter played a crucial role in seeding the growth of structure at an acceptable rate. A crucial ingredient to this picture is that the DM particles must have had a negligible kinetic energy at the time of structure formation.  If not, the formation of structures at large scales would be slowed down because DM particles with large kinetic energies would not cluster at sufficiently small scales due to free-streaming. This slowing down the growth of structures would wash out matter density fluctuations. It is therefore critical to investigate the suppression on the growth of structure caused by the fraction of DM which is non-thermally produced in a relativistic state via the aforementioned decay process, and to derive quantitative bounds on the fraction of DM particles produced in this manner.  

It has been shown that, at scales above the free-streaming length, relativistic DM particles effectively behave like cold DM. At small scales, the matter fluctuations of cold DM particles is governed by a linear equation which, during the matter-dominated stage of the Universe, follows the power-law behavior \cite{Ma}
\begin{equation}
\delta (f) \propto  a^{\alpha_{\infty}(f)}.
\label{delta}
\end{equation}
In the equation above, $\alpha_{\infty}(f)$ is the growth rate of the cold DM field, which during the matter-dominated epoch is given, to first order, by \cite{Ma}
\begin{equation}
\alpha_{\infty}(f) = \frac{5}{4}\sqrt{1-\frac{24}{25}f} \simeq 1 - 3/5 f,
\label{alpha}
\end{equation}
where $f$ is, as above, the fraction of the DM density produced non-thermally in a relativisitic state, and where $f\ll1$.
It is then possible to determine the suppression on small-scales caused by the non-thermal production of relativistic DM particles by comparing the matter fluctuation given in Eqs.(\ref{delta})-(\ref{alpha}) with the pure cold DM case at MRE. To first order, we find

\begin{equation}
g = \frac{\delta (f)}{\delta (f=0)} = a_{EQ}^{-3/5 f} \simeq \exp(-4.9 f).
\label{g}
\end{equation}
As mentioned above, it is important to notice that Eq.~(\ref{g}) is applicable only in the matter-dominated regime, and a more accurate calculation considering the effects of, e.g., a cosmological constant, is out of the scope of this work. 

After combining measurements of the amplitude of matter fluctuations in the Universe on scales of $8 h^{-1}$~Mpc from the WMAP9 results \cite{WMAP9} with Lyman-alpha forest data \cite{Lyman}, the resulting limit is $g > 0.95$. This limit implies, from Eq.~(\ref{g}), that $f < 0.01$.

In summary, structure formation constraints require that only a small fraction (less than $1\%$) of the DM in the Universe might have been produced in a relativistic state from a non-thermal process such as the decay we consider in the present framework.

\subsection{Big Bang Nucleosynthesis}

As mentioned above, the lifetime and the energy released by the mother particle $X^\prime$ are also constrained by BBN bounds.  This is because the energy released at a given time in the history of the Universe might induce electromagnetic showers that create and/or destroy light elements synthesized in the early universe \cite{BBN}. In this subsection we analyze the possible effects of this scenario on BBN.  

Given the overall quantitative success of BBN in the ``standard cosmological model'', the first requirement for new physics is that it not drastically alter any of the light elemental abundances such as those of D, ${}^4$He, ${}^7$Li. Energy releases which occur long after BBN may, in principle, spoil the successful BBN predictions for light
elemental abundances. The energy of photons produced in
late decays of $X^{\prime}$ would have been rapidly redistributed through scattering off
background photons ($\gamma \gamma_{BG} \rightarrow e^+ e^-$) as well as through inverse Compton scattering ($e\ \gamma \rightarrow e\ \gamma$) \cite{BBN,BBN2}. As a result, the constraints we obtain from BBN are, to an excellent approximation, independent of
the initial energy distribution of the injected photons and are only sensitive to the {\em total energy} released in the decay process \cite{BBN}.

In order to derive BBN limits on the fraction of relativistic, non-thermally produced DM via the decay $X^{\prime} \rightarrow DM + \gamma$, we therefore need to calculate the total electromagnetic energy released. 
Let $Y= n/n^{CMB}_{\gamma}$, where $n$ is the number density of particles of a particular species and $n^{CMB}_{\gamma}$ is the number density of CMB photons, given by
\begin{equation}
n^{CMB}_{\gamma}= \frac{2\zeta(2)}{\pi^2} T^3.
\end{equation}
The total electromagnetic energy released from the $X^\prime$ decay is thus $\varepsilon_{EM}= E_{\gamma} Y_{DM}$. If for each  $X^{\prime}$ particle we have the production of a DM particle plus a photon, then $Y_{X^{\prime}}=Y_{\gamma}=Y_{DM,\tau}=Y_{DM,0}$, where $Y_{DM,\tau}$ determines the number density of particles at a time equal to the lifetime of $X^{\prime}$, and $Y_{DM,0}$ is the number density of DM particles today. 

We thus find that the normalized number density of DM particles is given by
\begin{equation}
Y_{DM}=\frac{n_{DM}}{n^{CMB}_{\gamma}}= \frac{\Omega_{DM} \rho_c}{M_{DM} n^{CMB}_{\gamma,0}}.
\label{YDM1}
\end{equation}
This expression evaluates to
\begin{equation}
Y_{DM} \simeq 3\cdot10^{-14}\left( \frac{\rm TeV}{M_{DM}} \right) \left( \frac{\Omega_{DM}}{0.227} \right) \left( \frac{f}{0.01}\right).
\label{Yx}
\end{equation}

The factor $f$ showed up in  Eq.~(\ref{Yx}) because we assume here that only a fraction of the DM of the universe is produced in the decay process, while the majority of it is produced in a non-relativistically by some other mechanism that does not contribute any significant energy release at BBN (for example, via a standard thermal freeze-out process).

Since the photon energy produced in the decay is, 
\begin{equation}
E_{\gamma} =\frac{1}{ 2M_{ X^{\prime} } }(M^2_{X^{\prime}}-M^2_{DM}),
\end{equation}
the total electromagnetic energy released is given by
\begin{eqnarray}
\varepsilon_{EM} & = & 1.5\cdot10^{-11}\ \mbox{GeV}\times \nonumber\\
                 &   &
  \left( \frac{\Omega_{DM}}{0.227} \right) \left( \frac{f}{0.01}\right)  \left( \frac{M_{ X^{\prime} }}{M_{DM}} - \frac{M_{DM}}{M_{X^{\prime}} } \right).
\label{zeta}
\end{eqnarray}

In the limit $M_{X^{\prime}} \gg M_{DM}$ we can directly relate the total energy release given in Eq.~(\ref{zeta}) with the quantity $f\times\left(M_{ X^{\prime}}/M_{DM}\right)$ as well as with $\Delta N_{eff}$, as given in  Eq.~(\ref{deltaNeff}). 

\begin{figure}[!t]
\centering
\includegraphics[width=1\columnwidth]{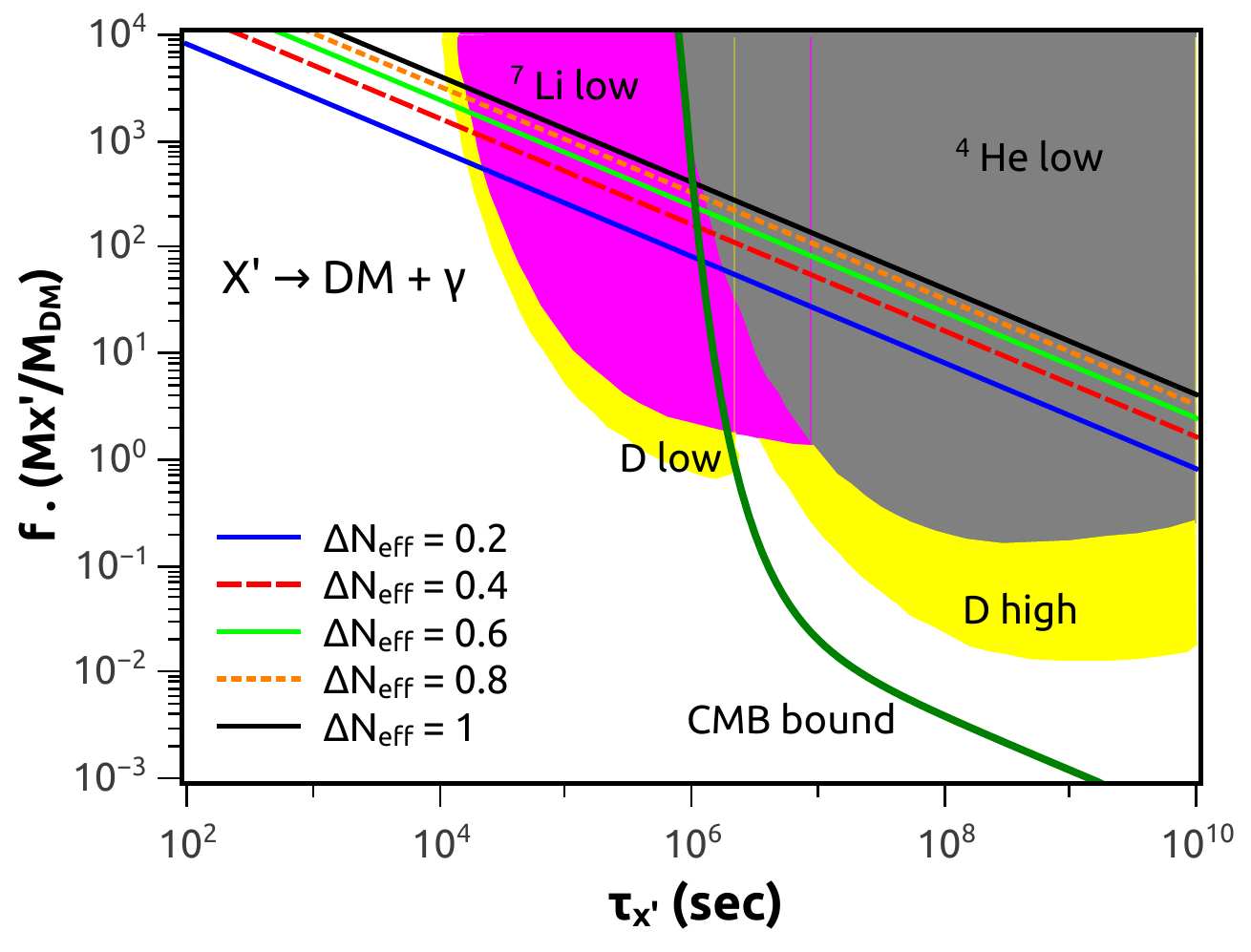}
\caption{The parameter space defined by the mother particle $X^\prime$ lifetime ($\tau_{X^\prime}$) and the fraction of relativistically produced DM ($f$) times the mother-to-daughter mass ratio $M_{X^\prime}/M_{DM}$, and constraints from BBN and CMB. The shaded regions show BBN bounds on the non-thermal production of DM via the decay $X^{\prime} \rightarrow DM + \gamma$, corresponding to an excess relativistic degrees of freedom $\Delta N_{eff}=0.2,0.4,0.6,0.8,1$. The green curve represents the CMB bound (regions to the right of the curve are ruled out). We assume $M_{X^{\prime}} \gg M_{DM}$.}
\label{Graph7}
\end{figure}

With this in mind, we can use the limits on the light elemental abundances derived in Ref.~\cite{BBN}. Since their results were obtained as a function of the total energy released, we straightforwardly convert them in terms of the quantity $f\times (M_{ X^{\prime}}/M_{DM})$. We show our results in Fig.~\ref{Graph7}, where the shaded regions show portions of parameter space ruled out by BBN, as calculated in Ref.~\cite{BBN} employing the ``baryometer'' parameter $\eta=n_b/n_{\gamma}=6\times 10^{-10}$. The straight lines indicate combinations of the quantity $f\times M_{ X^{\prime}}/M_{DM}$ and $\tau_{X^\prime}$ producing the $\Delta N_{eff}$ as in the labels.


Fig.~\ref{Graph7} shows that the BBN bounds are weak for early decays because at early times the
universe is hot and the initial photon spectrum is rapidly
thermalized, leaving just a few high-energy photons able to modify the light elemental abundances. However, for lifetimes longer than $10^4$~s, BBN excludes most of the relevant parameter space.

\subsection{Cosmic Microwave Background}

Similar to the BBN constraints, bounds from measurements of the CMB spectrum
are largely independent of the injected photon energy spectrum. These bounds also
depend primarily only on the total energy release.
The key effect of the additional energy injection in the
form of photons is related to spectral distortions to the CMB black-body spectrum \cite{CMBbound}.

For times $t  \lesssim 10^3$~s the processes of bremsstrahlung, i.e. $eX \rightarrow eX\gamma$ (where
X is an ion), Compton scattering and double Compton scattering $e \gamma \rightarrow e \gamma \gamma$  quickly thermalize the injected photon energy \cite{CMBbound}.
For $t  \gtrsim 10^3$~s, however, the bremsstrahlung and double Compton processes become inefficient, and the photon spectrum relaxes to a Bose-Einstein distribution with a chemical potential ($\mu$) different from zero. Limits on $\mu$ can then be used to constrain this additional injected energy component.

The change in the CMB spectrum caused by the energy injection is constrained by precise measurements from current satellites and telescopes. Following the procedure used in Ref.~\cite{CMBbound}, we calculated the limit (shown as a solid green curve in  Fig.~\ref{Graph7}) imposed by CMB data on the non-thermal production of DM at a given late time.  In particular, measurements of the CMB spectrum constrain any perturbation which might cause it to depart from an almost perfect blackbody spectrum.
These limits are expressed in terms of deviations of the chemical potential $\mu$ from 0. Current bounds imply that $\mu < 9 \times 10^{-5}$ \cite{cmb1, cmb2}. This upper limit can then be translated into a bound on the amount of energy released during the decay, providing a constrain on the quantity $f\times (M_{X^\prime}/M_{DM})$ for a given lifetime.

Fig.~\ref{Graph7} clearly shows that the CMB bounds become more restrictive than those from BBN for lifetimes $\tau_{X^\prime}$ longer than $\sim 10^6$~s, ruling out almost all the relevant parameter space to explain $\Delta N_{eff}$ with relativistic non-thermally produced WIMPs.

\section{FOUR ILLUSTRATIVE DARK MATTER MODELS}
\label{sec:models}

To summarize, we found thus far that constraints from structure formation, BBN and CMB require (i) $f < 1\%$ and (ii) a lifetime for the mother particle $X^\prime$  shorter than $10^4$~s. For concreteness, we present here four example dark matter models, including a supersymmetric case, where DM particles could be partly produced non-thermally in association with a photon. We also show that these models have the potential to reproduce $\Delta N_{eff} \simeq 1$ while obeying the cosmological bounds described in the previous section. Unless otherwise noted, we shall use the symbol $M_{X^\prime}$ for the mass of the mother particle and $M_{DM}$ for the mass of the DM particle, while we use different symbols to indicate the associated fields.

Before discussing specific models, we note that, in general, we can recast Eq.~(\ref{deltaNeff}) as
\begin{equation}
\left(\frac{\tau}{10^4\ {\rm s}}\right)^{1/2}\simeq4\times 10^5\ \Delta N_{eff}\left(\frac{0.01}{f}\right)\frac{M_{DM} M_{X^\prime}}{\left(M_{X^\prime}-M_{DM}\right)^2}.
\end{equation}
This relation, in view of the constraints obtained above, namely that $\tau\gtrsim10^4$ s and that $f\lesssim 0.01$, implies
\begin{equation}
\label{eq:ratio}
\frac{M_{X^\prime}}{M_{DM}}\gtrsim4\times 10^5\ \Delta N_{eff}.
\end{equation}
We thus find that the ratio between the mother and the DM particle, for appreciable values of $\Delta N_{eff}$, must be at least of the order of $10^5$. We then conclude that to produce a large enough number of effective relativistic degrees of freedom, the mother particle and the DM particle {\em cannot be nearly degenerate}. In particular, this rules out scenarios where the two relevant particles are expected to be close in mass, such as, for example, in any universal extra dimensional setup (see e.g. Ref.~\cite{fengetal}).

\subsection{Spin-0 particle}\label{sec:spin0}
In this first case, the mother particle $X^\prime$ is a heavy scalar ($S$, with mass $M_{X^\prime}$) which decays into a spin-1, stable DM particle ($B_\mu$, with mass $M_{DM}$) plus a photon. The relevant Lagrangian interaction term is given by the effective dimension-5 operator

\begin{equation}\label{eq:leffspin0}
\mathcal{L}_{eff}= \frac{1}{\Lambda} B_{\mu} (\partial_{\nu} S)   F^{\mu \nu},
\label{eq:spin-0}
\end{equation}where $F^{\mu \nu}$ is the $U(1)$ electromagnetic field tensor.  

The resulting lifetime of the mother particle in this case is found to be \footnote{As explained above, we are well in the limit where the mother particle is much heavier than the DM particle, see Eq.~(\ref{eq:ratio}); we thus neglect $M_{DM}/M_{X^\prime}$ factors in the expression for the lifetime.}

\begin{equation}\label{eq:spin0lifetime}
\tau (S \rightarrow \gamma B) = \left(1.32\times10^{-22}\ {\rm s}\right)\left(\frac{\Lambda}{M_{X^\prime}}\right)^2\left(\frac{\rm GeV}{M_{X^\prime}}\right),
\end{equation}
with $\Lambda$ the cutoff scale of the effective theory, as in Eq.~(\ref{eq:leffspin0}). 

\begin{figure}[!t]
\centering
\includegraphics[width=1\columnwidth]{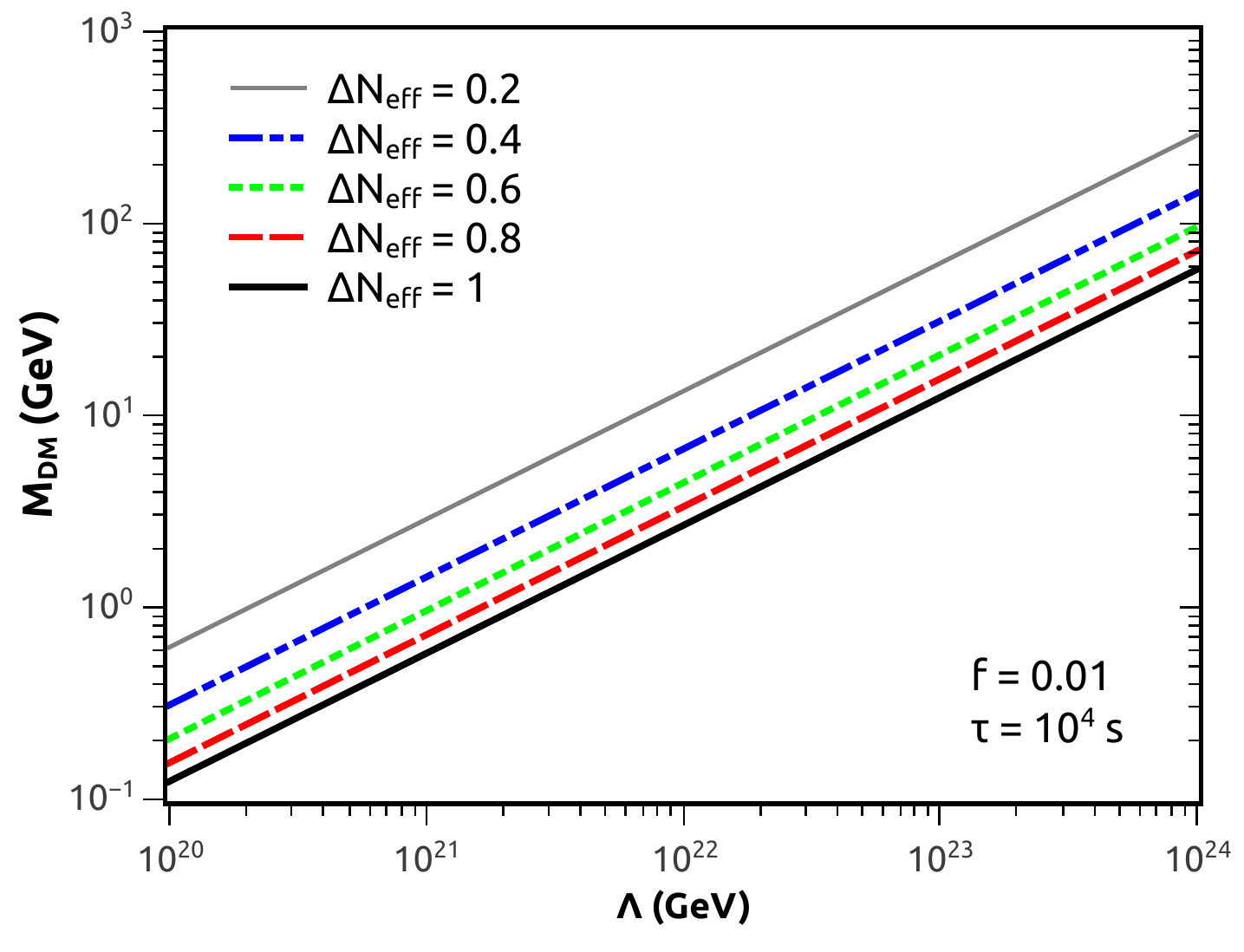}
\caption{Lower limits on the DM mass $M_{DM}$ for a model in which a heavy scalar particle decays into a spin-1 DM particle plus a photon, for different $\Delta N_{eff}$. The relevant effective operator for this model is given in Eq.~(\ref{eq:spin-0}).  We assume in this figure $f=0.01$ and a lifetime $\tau=10^4\,$s. }
\label{fig:spin-0}
\end{figure}

The upper limit on the lifetime we derived above, $\tau<10^4\ {\rm s}$,  can then be recast as a lower limit on the mother particle mass, for a given cutoff scale $\Lambda$.  This lower limit on the mother particle mass can then also be, in turn, converted to a lower limit on the DM particle mass $M_{DM}$, once values for the parameters $f$ and $\Delta N_{eff}$ are fixed.  Using the expression for the lifetime, along with  Eq.~(\ref{deltaNeff}), we obtain the requirement 
\begin{equation}
M_{DM}\lesssim\left(5.75\cdot10^{-13}{\ \rm GeV^{1/3}}\right)\left(\frac{\tau}{10^4\ {\rm s}}\right)^{1/6}\Lambda^{2/3}\frac{f}{\Delta N_{eff}}.
\label{eq:mdm1}
\end{equation} 

In Fig.~\ref{fig:spin-0} we plot the lower limit for the DM mass as a function of the cutoff scale $\Lambda$, for $f=0.01$ and $\tau=10^4\,$s. To obtain DM masses in the $10$~GeV mass range in this effective model would require $\Lambda \sim 10^{22}$~GeV.  Had we included a coupling constant $g$ in Eq.~(\ref{eq:spin-0}), we would have concluded that for $M_{DM}\sim 10$ GeV, if $\Lambda$ is of the order of the GUT scale ($\Lambda \sim10^{16}$~GeV) the required coupling would have been $g\sim10^{-3}$. 

\subsection{Spin-1 particle}

\begin{figure}[!t]
\centering
\includegraphics[width=1\columnwidth]{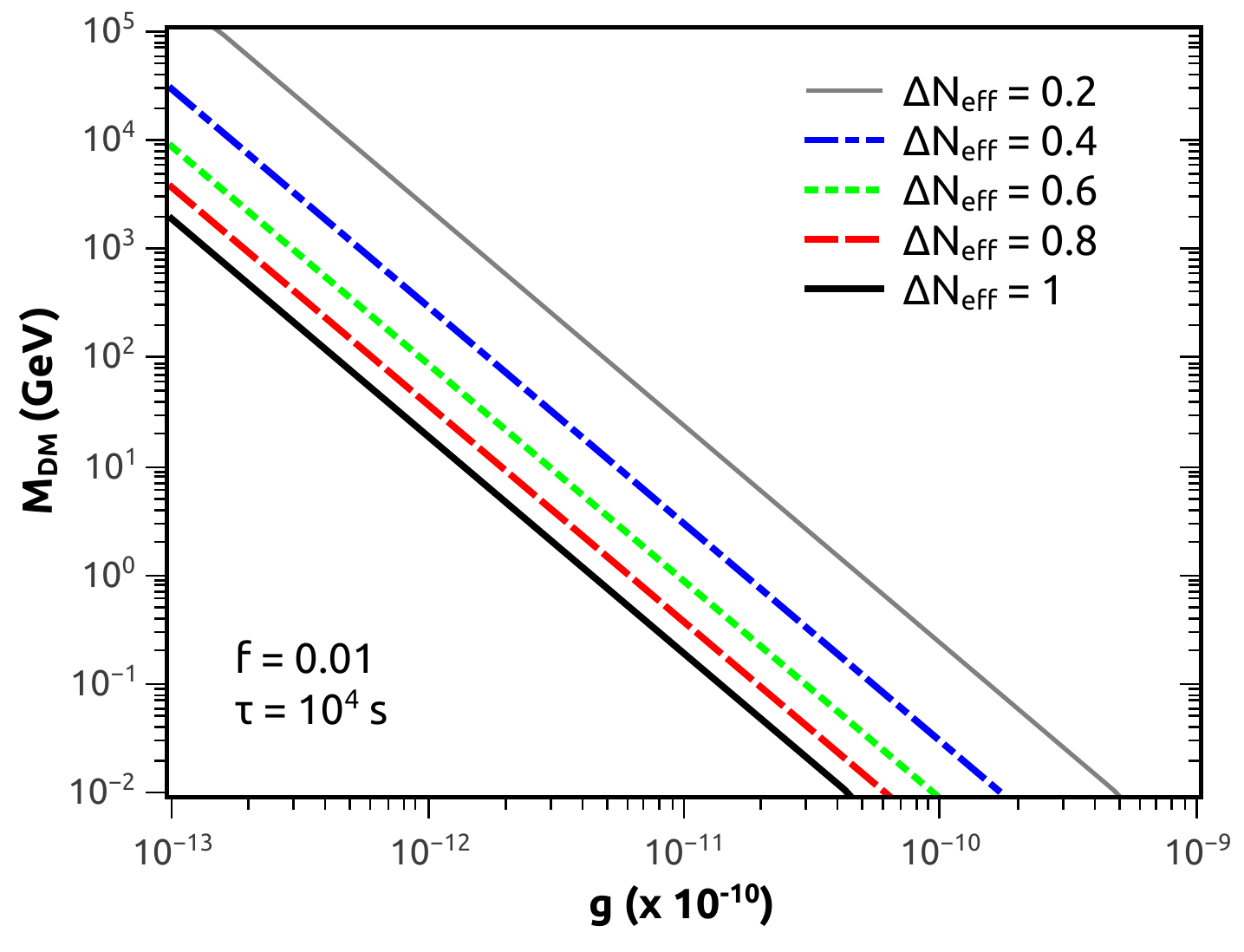}
\caption{Lower limits on the DM mass for a model in which a heavy spin-1 boson decays into a spin-1 DM particle plus a photon, for different $\Delta N_{eff}$. The relevant operator for this effective model is given in Eq.~(\ref{eq:spin-1}).  We employed $f=0.01$ and a lifetime $\tau=10^4\,$s. Note that the x-axis is multiplied by a factor  $10^{-10}$. }
\label{fig:spin-1}
\end{figure}

In this case, a heavy spin-1 boson ($W_\mu$, again with mass $M_{X^\prime}$) decays into a spin-1 DM particle ($B_\mu$, of mass $M_{DM}$) plus a photon. A Lagrangian term which mediates this decay through a renormalizable operator is

\begin{equation}
\mathcal{L}_{eff}= g W_{\mu} B_{\nu} F^{\mu \nu},
\label{eq:spin-1}
\end{equation}
where, again, $F^{\mu \nu}$ is the Standard Model electromagnetism field tensor. The lifetime of the mother particle in this case is
\begin{equation}
\tau = \frac{2.65\cdot10^{-22}{\ \rm s}}{g^2}\left(\frac{M_{DM}}{M_{X^\prime}}\right)^2\left(\frac{\rm GeV}{M_{X^\prime}}\right).
\end{equation}
We can again use the upper limit on the lifetime $\tau\gtrsim10^4\ {\rm s}$ and Eq.~(\ref{deltaNeff}) to set a lower limit on the DM mass, 
\begin{equation}
M_{DM}\lesssim\frac{2.93\cdot10^{-36}{\rm \,GeV}}{g^2}\left(\frac{\tau}{10^4\ {\rm \,s}}\right)^{1/2}\left(\frac{f}{\Delta N_{eff}^{\nu}}\right)^3.
\label{eq:mdm2}
\end{equation} 
We can thus calculate the values for $g$ and for the DM particle mass producing a given number of effective relativistic degrees of freedom $\Delta N_{eff}$, once the fraction of DM produced non-thermally ($f = 0.01$) and the lifetime ($\tau=10^4$ s) are set to values compatible with BBN and Structure Formation bounds. We show our results in Fig.~\ref{fig:spin-1}.

The figure implies that in this particular scenario a DM particle with a mass in the $\sim 10$~GeV range might account for a $\Delta N_{eff}\sim1$ extra radiation component for $g \sim 10^{-22}$.

\subsection{Spin-1/2 particle}

In this case a heavy fermion ($\psi$, with mass $M_{X^\prime}$) decays into a spin-1/2 DM particle ($\chi$, mass $M_{DM}$) plus a photon according to the effective dimension-5 operator

\begin{equation}
\mathcal{L}_{eff}=\frac{1}{\Lambda} \left(\bar{\psi}\ \sigma_{\mu \nu}\ \chi\right) F^{\mu \nu}+{\rm h.c.}
\label{Lfermion}
\end{equation}
The lifetime of the mother particle in this scenario is
\begin{equation}
\tau = \left(4.14 \times 10^{-24}{\ \rm s}\right)\left(\frac{\Lambda}{M_{\psi}}\right)^2\left(\frac{\rm GeV}{M_{\psi}}\right)
\end{equation}

\begin{figure}[!t]
\centering
\includegraphics[width=1\columnwidth]{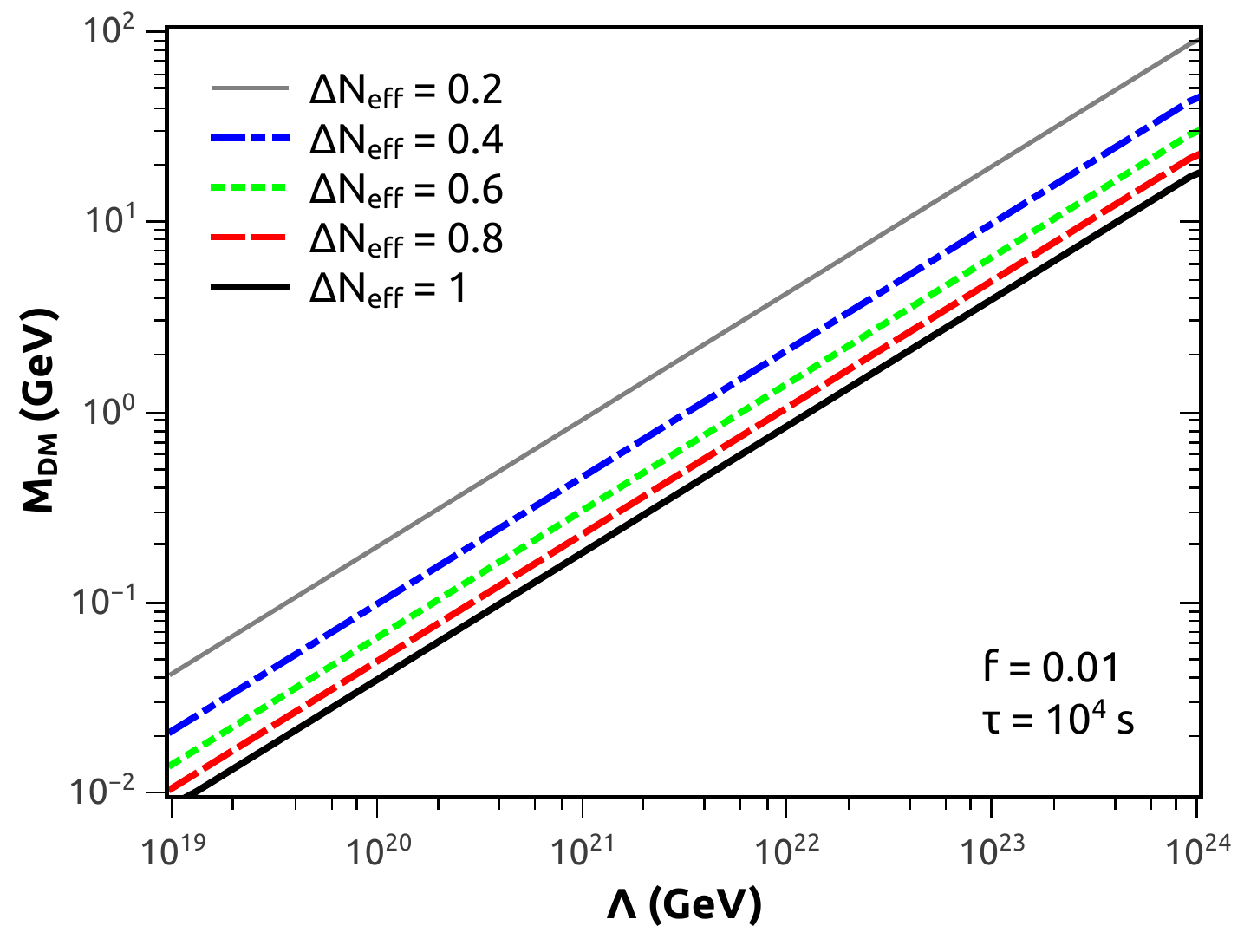}
\caption{Lower limits on the DM mass for a model in which a heavy fermion decays into a spin-1/2 DM particle plus a photon, for different $\Delta N_{eff}$. The relevant operator for this effective model is given in Eq.~(\ref{Lfermion}).  We employed $f=0.01$ and a lifetime $\tau=10^4\,$s.}
\label{fig:spin-12}
\end{figure}

We can again use the upper limit on the lifetime to set a lower limit on the DM particle mass, 

\begin{equation}
M_{DM}\lesssim\left(1.81\cdot10^{-13}{\ \rm GeV^{1/3}}\right)\left(\frac{\tau}{10^4\ {\rm \,s}}\right)^{1/6}\Lambda^{2/3}\left(\frac{f}{\Delta N_{eff}}\right).
\label{eq:mdm12}
\end{equation} 
This case closely mirrors the spin-0 example of Sec.~\ref{sec:spin0}, the only difference being a numerical factor of 32 in the lifetime.  This translates to moving the DM mass upper limit down by a factor of $32^{1/3}\approx 3$. Therefore we conclude that, similarly to the spin-0 case, the non-thermal production of a $\sim 10$~GeV DM particle can mimic one extra relativistic neutrino in the early universe for $\Lambda \sim 10^{22}$~GeV, as shown in Fig.~\ref{fig:spin-12}.

\subsection{A Supersymmetric Example: the Bino/Gravitino System}

Neutralinos are the mass eigenstates resulting from a mixture of  neutral B-ino, W-ino, and Higgs-inos. In many supersymmetric models, including certain supergravity scenarios with universal soft supersymmetry breaking masses at the grand unification scale, the lightest neutralino is an almost pure Bino. The lightest neutralino is also the next-to-lightest supersymemtric particle with the lightest supersymmetric particle being the gravitino. Binos decay into a gravitino-photon final state via the interaction Lagrangian term  \cite{Jfeng1}
\begin{equation}
\mathcal{L}= \frac{-i}{8\pi M_{\star}} \tilde{G_{\mu}}[\gamma^{\nu},\gamma^{\rho}]\gamma^{\mu}\tilde{B^{\mu}}F_{\nu \rho},
\label{eqgravitino}
\end{equation}where $M_{\star}=2.4\times 10^{18}$~GeV is the reduced Planck mass. 
In models with low-scale supersymmetry breaking $M_{\tilde{B}} \gg M_{\tilde{G}}$, which is exactly the limit needed to realize the scenario we are interested in. In this case, from Eq.~(\ref{eqgravitino}) we find a neutralino lifetime of
\begin{equation}\label{eq:binolifetime}
\tau (\tilde{B} \rightarrow \gamma  \tilde{G}) \simeq 750\ {\rm s}\left(\frac{M_{\tilde G}}{1\ \rm keV}\right)^2\left(\frac{1\ {\rm GeV}}{M_{\tilde B}}\right)^5.
\end{equation}
Similarly to previous sections, we can convert the bound on the lifetime into a lower
limit on the DM mass,
\begin{figure}[!t]
\centering
\includegraphics[width=1\columnwidth]{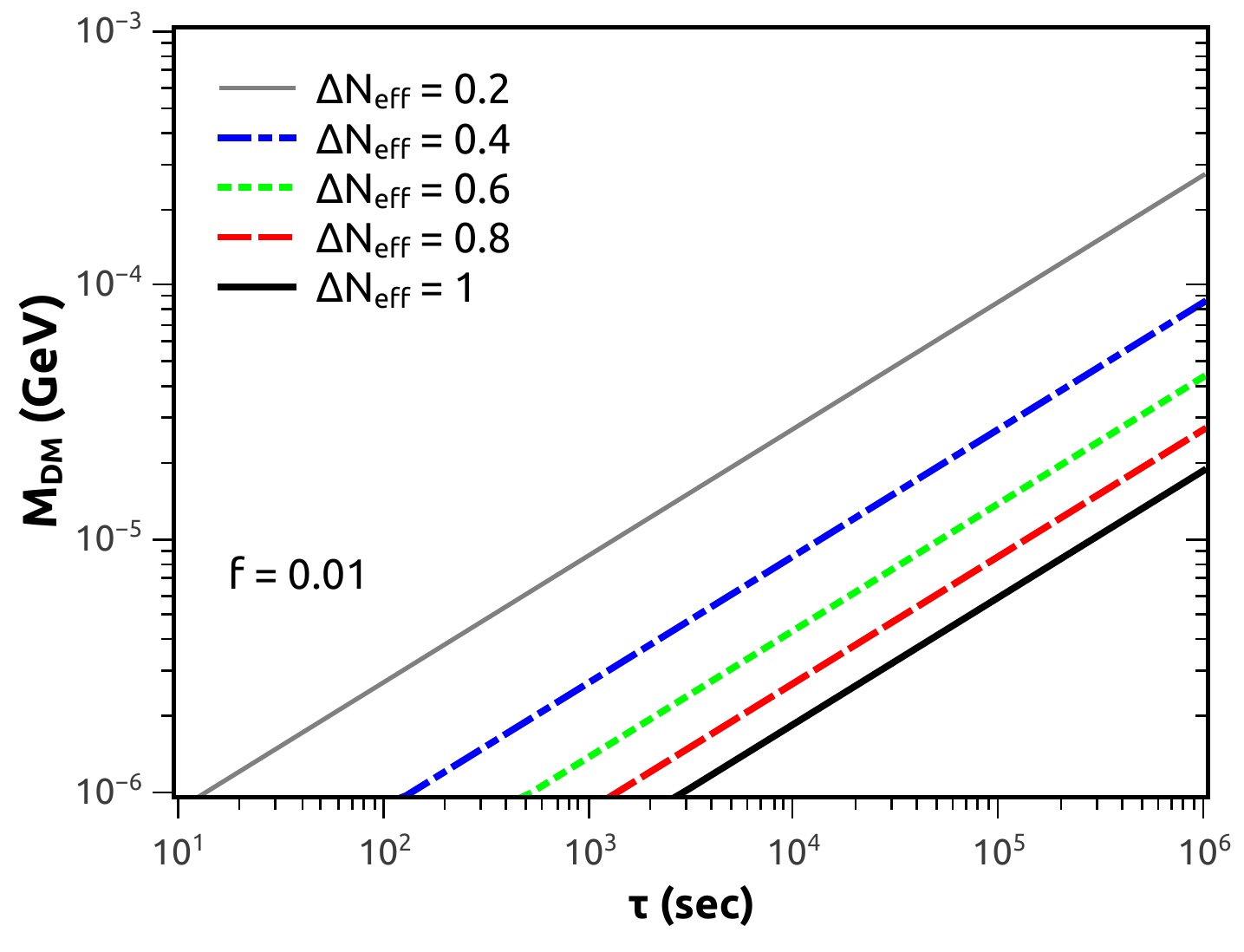}
\caption{Lower limit on the gravitino mass, for a supersymmetric model where a fraction $f$ of gravitinos are non-thermally produced by the decay of pure Binos. The relevant lifetime is derived from Eq.~(\ref{eqgravitino}). As in the previous figures, we have used $f=0.01$ and . }
\label{figravitino}
\end{figure}
\begin{equation}
M_{\tilde{G} } \lesssim (4\ {\rm MeV}) \left( \frac{\tau}{10^4}\right)^{1/2} \left( \frac{f}{\Delta N_{eff}}\right)^{5/3}.
\end{equation}
Setting, as above, $f=0.01$ and $\tau=10^4$ s, we show in Fig.~\ref{figravitino} that a gravitino with mass $M_{\tilde G}=M_{DM}$ in the 2 to 20 keV range mimics the effect of an extra neutrino while still obeying cosmological bounds.  We note that such light gravitinos are marginally consistent with bounds from small scale structure (for example Lyman-alpha forest data, see e.g. Ref.~\cite{Viel:2005qj}). Also, assuming a gluino mass in the TeV range, gravitinos in the mass range needed here would require a reheating temperature $T_R$ close to the electro-weak scale to avoid over-closing the universe, since the thermal production of gravitinos in the early universe  \cite{Bolz:2000fu} implies
\begin{equation}
\frac{T_R}{100\ {\rm GeV}}\simeq \left(\frac{\Omega_{\tilde G}h^2}{0.2}\right)\left(\frac{1\ {\rm keV}}{M{\tilde G}}\right).
\end{equation}
Such a low reheating temperature would rule out certain scenarios for the production of the baryon asymmetry in the universe, such as leptogenesis, but is in general not phenomenologically implausible. 

Finally, we note that the constraints on the lifetime and Eq.~(\ref{eq:binolifetime}) imply a lower limit on the bino mass, which must be larger than about 1 GeV.  This is, of course, perfectly compatible with the supersymmetric models relevant here.

\section{CONCLUSIONS}
\label{sec:conclusion}
Recent data show tentative evidence for excess effective radiation degrees of freedom in the early universe. In this study, we analyzed scenarios where such extra radiation stems from a sub-dominant population of WIMP DM particles which is produced relativistically from the decay of a heavier particle. Such decay could affect structure formation, the cosmic microwave background, and the production of light elements in the early universe. This framework is therefore constrained in a model-independent way by various bounds from cosmological observations.

We showed that structure formation limits the {\em fraction} of the DM energy density that can be produced in the aforementioned relativistic state to less than a percent, while CMB and BBN force the decay when the DM is produced to occur with a lifetime shorter than about $10^4$ s. Given these two key constraints, we showed that a large enough contribution to the effective relativistic degrees of freedom $\Delta N_{eff}$ is achieved only for a relatively large mass hierarchy (on the order of $10^5$) between the mother ($X^\prime$) and daughter ($DM$) particle masses in the decay (for example, $X^\prime\to DM+\gamma$).

We presented a set of illustrative particle DM models that implement this scenario, including effective theories as well as a supersymmetric example. For each example, we derived the relevant combination of parameters leading to a significant $\Delta N_{eff}$, while being consistent with all cosmological bounds from structure formation, BBN and CMB.

As a final remark, we note that in the context of the present framework it is in principle easy to accommodate both:
\begin{itemize}
\item[(i)] $M_{DM} \simeq 10$~GeV, i.e. in the mass region of particular interest today from both direct detection \cite{DDsignals} and indirect detection \cite{GCexcess}, and which has been extensively investigated in many particle physics models \cite{DDmodels1,DDmodels2,DDmodels3}, or 
\item[(ii)] assume that the DM particle mass is in the classic WIMP $\sim 100$~GeV range, which is also particularly interesting from the standpoint of possible indirect detection signals \cite{IDsignals1,IDsignals2,IDsignals3,GCexcess}, and which has also been exhaustively explored in particle physics models as well, see e.g. Ref. \cite{IDmodels1,IDmodels2,IDmodels3,IDmodels4}.
\end{itemize}

For any DM particle mass, the key requirements that the present setup imposes are to have:
\begin{enumerate} 
\item a large enough mother particle mass $M_{X^{\prime}}$ compared to the dark matter mass, to produce a sizable $\Delta N_{eff}$ according to Eq.~(\ref{deltaNeff}),
\item a short enough lifetime for decay into the DM particle ($\tau < 10^4$~s) to evade BBN and CMB constraints, and 
\item a small enough fraction of relativistically produced particles ($f\lesssim 0.01$) to evade structure formation bounds.
\end{enumerate}

\acknowledgments
This work is partly supported by  the Department of
Energy under contract DE-FG02-04ER41286 (SP), and by the Brazilian National Counsel for Technological and
Scientific Development (CNPq) (FQ).

\appendix

\section{Boost Factor}
\label{boostfactor}

The boost factor of a DM particle produced in the decay $X^{\prime} \rightarrow DM + \gamma$ can be found using energy and momentum  conservation,
\begin{eqnarray}
\label{daneq1}E_{X^{\prime}} &=& E_{\gamma} + M_{DM}\ \gamma_{DM},\\
P_{X^{\prime}} &=& E_{\gamma} + P_{DM}.
\label{daneq2}
\end{eqnarray}
Assuming that $X^{\prime}$ decays at rest, we find
\begin{equation}
| P_{DM} | = | E_{\gamma} |=  \frac{1}{ 2M_{ X^{\prime} } }(M^2_{X^{\prime}}-M^2_{DM}).
\label{daneq3}
\end{equation}
Substituting Eq.~(\ref{daneq3}) into Eq.~(\ref{daneq1}) and taking $E_{X^{\prime}}=M_{X^{\prime}}$ we get
\begin{equation}
\gamma_{DM}=  \left( \frac{ M_{X^{\prime}} }{2M_{DM}} + \frac{M_{DM}}{2M_X^{\prime}} \right).
\end{equation}

\noindent Taking into account the expansion of the universe and assuming that all decays happen at the decay lifetime $t=\tau$, we find
\begin{equation}
\gamma_{DM}= 1+ \left( \frac{a_{\tau}}{a} \right) \left[\left( \frac{ M_{X^{\prime}} }{2M_{DM}} + \frac{M_{DM}}{2M_X^{\prime}} -1 \right) \right] .
\end{equation}

\noindent At the time of decay, the universe was dominated by radiation with $a=(t/t_0)^{1/2}$, thus
\begin{equation}
\gamma_{DM}= 1+ \left( \frac{\tau}{t} \right)^{1/2} \left[\left( \frac{ M_{X^{\prime}} }{2M_{DM}} + \frac{M_{DM}}{2M_X^{\prime}} -1 \right) \right] .
\label{gammaX}
\end{equation}
Therefore, at matter-radiation equality ($t=t_{EQ}$) we obtain
\begin{equation}
\gamma_{DM} \simeq 1+ 7.8\times 10^{-4}\left( \frac{\tau}{10^6 s} \right)^{1/2} \left[\left( \frac{ M_{X^{\prime}} }{2M_{DM}} + \frac{M_{DM}}{2M_X^{\prime}} -1 \right) \right] .
\label{gammaX2}
\end{equation}
Eq.~(\ref{gammaX2}) gives us the boost factor of the DM particle as a function of time and of the mother-to-daughter mass ratio, for the case where all decays happen when the universe was radiation dominated. 

Since the additional radiation density reads
\begin{equation}
\rho_{\mbox{extra}}= f \times \rho_{DM} (\gamma_{DM} -1),
\label{rhoextraEq}
\end{equation}where $f$ is the fraction of DM non-thermally produced in the decays, we conclude that the number of effective relativistic degrees of freedom at matter-radiation equality associated with the non-thermal production of DM is $\Delta N_{eff} = \rho_{extra}/\rho_{1 \nu} $, where $\rho_{1 \nu}$ is the number density of one neutrino species at the same epoch. Using Eq.~(\ref{rhoDMnu}), we find
\begin{equation}
\Delta N_{eff} = \frac{f (\gamma_{DM} -1)}{0.16}.
\label{Neffeq}
\end{equation}
Thus, substituting Eq.~(\ref{gammaX}) into Eq.~(\ref{Neffeq}), we obtain
\begin{eqnarray}
\Delta N_{eff} & \simeq & 4.87 \times 10^{-3}\left( \frac{\tau}{10^6\ s} \right)^{1/2}\nonumber\\ 
                     &   &
  \times \left[\left( \frac{ M_{X^{\prime}} }{2M_{DM}} + \frac{M_{DM}}{2M_X^{\prime}} -1 \right) \right]\times f.
\label{deltaNeff2}
\end{eqnarray}
This expression indicates that if some fraction of the DM of the universe is produced non-thermally via the decay $X^{\prime} \rightarrow DM + \gamma$, for reasonable values of the lifetime and mass ratio ($M_{X^{\prime}}/M_{DM}$), this production mechanism produces an effective excess of relativistic degrees of freedom.

\end{document}